\documentclass[conference]{IEEEtran}
\IEEEoverridecommandlockouts
\usepackage{cite}
\usepackage{listings}
\usepackage{amsmath,amssymb,amsfonts}
\usepackage{graphicx}
\usepackage{textcomp}
\usepackage{xcolor}
\usepackage{amsmath} 
\usepackage{booktabs} 
\usepackage{multirow} 
\usepackage{array} 
\usepackage{float}
\usepackage{placeins}
\usepackage{hyperref}
\usepackage{algorithm}
\usepackage{algpseudocode}
\usepackage{amsmath} % For \max, \text
\usepackage{amssymb} % For \emptyset
\usepackage{siunitx}
\usepackage{tabularx}
\usepackage[font=normalsize]{caption}
\usepackage[belowskip=2pt,aboveskip=2pt]{caption}
\definecolor{backgroundColour}{HTML}{F8F8F8}
\definecolor{keywordclr}{HTML}{AA00FF}
\definecolor{commentclr}{HTML}{649696}
\definecolor{stringsclr}{HTML}{11A579}
\definecolor{fnctionclr}{rgb}{0.467, 0, 0.533}
\definecolor{builtinclr}{rgb}{0.35, 0, 0.533}
\definecolor{symbolsclr}{rgb}{0.5, 0.25, 0.25}   
\definecolor{numbersclr}{rgb}{0.8, 0.2, 0}
\definecolor{bckgrndclr}{rgb}{0.91, 0.95, 0.95}
\lstdefinestyle{PythonStyle}{
    language=Python,
    backgroundcolor=\color{backgroundColour},
    keywordstyle=\color{keywordclr},
    stringstyle=\color{stringsclr},
    commentstyle=\color{commentclr},
    upquote=true,
    basicstyle=\ttfamily\linespread{0.9}\footnotesize,
    breakatwhitespace=false,
    breaklines=true,
    captionpos=b,
    keepspaces=true,
    numbers=left,
    numbersep=5pt,
    numberstyle=\color{commentclr}\ttfamily\tiny,
    showspaces=false,
    showstringspaces=false,
    showtabs=false,
    tabsize=2,
    xleftmargin=2.25em,
    frame=single,
    framexleftmargin=1.25em,
    morekeywords={assert,with,as,None}
}

\def\BibTeX{{\rm B\kern-.05em{\sc i\kern-.025em b}\kern-.08em
    T\kern-.1667em\lower.7ex\hbox{E}\kern-.125emX}}
\begin{document}

% \title{DRAwFed: Data Readiness Aware Federated Learning}
\title{FedCostAware: Enabling Cost-Aware Federated Learning on the Cloud}

\author{\IEEEauthorblockN{
Aditya Sinha\IEEEauthorrefmark{1}\IEEEauthorrefmark{2}\IEEEauthorrefmark{3},
Zilinghan Li\IEEEauthorrefmark{1},
Tingkai Liu\IEEEauthorrefmark{2}\IEEEauthorrefmark{3},
Volodymyr Kindratenko\IEEEauthorrefmark{2}\IEEEauthorrefmark{3},
Kibaek Kim\IEEEauthorrefmark{4},
Ravi Madduri\IEEEauthorrefmark{1}
}
\IEEEauthorblockA{
\IEEEauthorrefmark{1}Data Science and Learning Division, Argonne National Laboratory \\
\IEEEauthorrefmark{2}The Grainger College of Engineering, University of Illinois at Urbana-Champaign \\
\IEEEauthorrefmark{3}Center for AI Innovation, National Center for Supercomputing Applications \\
\IEEEauthorrefmark{4}Mathematics and Computer Science Division, Argonne National Laboratory \\
}
\IEEEauthorblockA{
\{aditya47, tingkai2, kindrtnk\}@illinois.edu, \{zilinghan.li, kimk, madduri\}@anl.gov
}
}

% \author{\IEEEauthorblockN{Aditya Sinha}
% \IEEEauthorblockA{\textit{University of Illinois at Urbana-Champaign} \\
% USA \\
% hiniduma.1@osu.edu}
% \and
% \and
% \and

\maketitle

\begin{abstract}
Federated learning (FL) is a distributed machine learning (ML) approach that allows multiple clients to collaboratively train ML models without exchanging original training data, offering a solution that is particularly valuable in sensitive domains such as biomedicine. However, training robust FL models often requires substantial computing resources from participating clients, which may not be readily available at institutions such as hospitals. While cloud platforms offer on-demand access to such resources, their usage can incur significant costs, particularly in distributed training scenarios where poor coordination strategies can lead to substantial resource wastage. To address this, we introduce FedCostAware, a cost-aware scheduling algorithm designed to optimize synchronous FL on cloud spot instances. FedCostAware addresses the challenges of training on spot instances and different client budgets by employing intelligent management of the lifecycle of spot instances. This approach minimizes resource idle time and overall expenses. Comprehensive experiments across multiple datasets demonstrate that FedCostAware significantly reduces cloud computing costs compared to conventional spot and on-demand schemes, enhancing the accessibility and affordability of FL.
\end{abstract}

\begin{IEEEkeywords}
Federated Learning, Distributed Computing, Cloud Computing, Cost Optimization, Scheduling Algorithm, Spot Instances, Budget Constraints
\end{IEEEkeywords}

\section{Introduction}
\label{sec:introduction}
Federated learning (FL) \cite{mcmahan2017communication,li2020federated,kim2024privacy} has emerged as a prominent collaborative machine learning paradigm, allowing multiple clients to train a shared model without exchanging raw data. FL proceeds by performing two key steps iteratively until a specified stopping criterion is met: (1) each client uses its own computing resources to perform local training on its private dataset and transmits the resulting model updates to a central server; (2) the server aggregates the client local models to update the global model and then distributes the updated model back to the clients. As no training data leaves each client's local machine, the privacy of the client's local data becomes easier to preserve. This data-decentralized approach is particularly valuable in sensitive domains such as biomedicine \cite{fl_medimg, fl_cancer, hoang2023enabling}, where regulatory constraints often restrict data sharing needed for centralized data collection. Therefore, FL provides a practical approach to developing privacy-preserving models in collaborating hospitals, laboratories, and research institutions. However, as with many contemporary ML endeavors, training effective models in these domains often requires high-performance computing resources, particularly access to GPUs \cite{pfitzner2021federated}. Maintaining a dedicated GPU infrastructure involves significant capital investment and ongoing operational costs, which can be cost-prohibitive for some organizations. Cloud computing presents a more flexible alternative by offering GPU instances on a pay-as-you-go basis, typically billed hourly. This allows researchers to provision computational power precisely when needed, aligning costs with usage. However, prices for standard on-demand instances can still be a substantial expense, especially for lengthy training periods. Consequently, cloud providers offer spot instances - the unused compute capacity available at considerably lower prices. Although spot instances risk preemptive termination, ML workflows can adapt using techniques like regular checkpointing to effectively utilize these volatile yet lower-cost resources. These capabilities position spot instances as an increasingly attractive and cost-efficient option for researchers operating under budget constraints.

In addition to the cost of model training itself, the key challenge of FL---the heterogeneity of client data and computing capabilities \cite{ye2023heterogeneous}---might introduce additional costs in FL settings. In typical real-world deployments, institutions such as major hospitals may have large datasets and powerful computing resources, while smaller clinics may possess smaller data and less performant machines. In synchronous FL algorithms, this imbalance in data amount and compute power causes the straggler problem \cite{ye2023heterogeneous}, where the overall training pace is dictated by the slower participants. Faster clients are forced to idle while awaiting straggler completion before global aggregation can occur. This inefficiency is particularly acute in cloud environments where costs accrue per GPU hour, even during idle time. The delays induced by stragglers translate directly into increased expenditures and prolonged training durations. Accumulated over numerous communication rounds, these delays significantly impede the cost-effectiveness of the FL training. Reducing the additional costs caused by the straggler problem is therefore an important challenge, especially for deployments leveraging variable-cost resources like spot instances under strict budgets.

Some approaches, notably asynchronous aggregation strategies, have been explored in the literature to reduce client idle time in FL \cite{xie2019asynchronous,nguyen2022federated,li2024fedcompass,iakovidou2024asynchronous}, but they present significant trade-offs. In asynchronous FL algorithms, the server updates the global model immediately upon receiving one or a few client local models and returns the updated model to clients for further local training. This asynchronous paradigm can potentially improve system throughput by eliminating or significantly reducing client waiting times. However, the frequent update of the global model commonly introduces the challenge of staleness of updates, where local gradients computed using outdated global models are incorporated by the central server. Previous comparative studies \cite{wilhelmi2022analysisevaluationsynchronousasynchronous, dun2023efficient} indicate that the temporal inconsistency inherent in asynchronous FL typically results in lower accuracy of the final model and degraded performance compared to synchronous training. Such potential degradation raises critical concerns, particularly in precision-sensitive fields such as biomedicine, where model accuracy and stability are paramount \cite{hoang2023enabling}. Alternatively, this work adopts the synchronous update protocol, which ensures consistency by utilizing updates computed from the identical global model state during the aggregation. However, as established, this adherence to synchronicity renders the training process vulnerable to straggler-induced inefficiencies and the associated computational costs, posing a key limitation for cost-effective deployment, especially in cloud environments.

% Although FL on spot instances offers cost benefits, existing approaches often overlook the combined challenge of optimizing dynamic spot costs within synchronous protocols while adhering to heterogeneous participant budgets. To address this gap, 
To address the above challenges, in this paper, we introduce FedCostAware, a cost-aware scheduling algorithm for synchronous FL using spot instances on public cloud computing resources, specifically designed for optimizing dynamic spot costs within synchronous protocols while adhering to different participant budgets. Our novel scheduler achieves this by performing several key functions:

\begin{itemize}
    \item \textbf{Intelligent Lifecycle Management:} Manages spot instances lifecycle by terminating them during idle periods (for example, while waiting for stragglers) and preemptively restarting them for subsequent training rounds to minimize costs.

    \item \textbf{Dynamic Cost Optimization:} Selects the most economical cloud regions and availability zones for spot instance provisioning in real-time based on current pricing.
    
    \item \textbf{Enabling Client Budgets:} Allows individual clients to specify their maximum budget for the FL experiment.
    
    \item \textbf{Fault-Tolerant Spot Utilization:} Checkpoints weights and adjusted schedules to recover from interruptions, minimizing lost work and idle GPU costs.
    
\end{itemize}

Experiments on several naturally and synthetically partitioned federated datasets indicate that leveraging spot instances can reduce FL training costs by approximately 60\% compared to on-demand instances. Moreover, incorporating the proposed FedCostAware scheduling algorithm can further improve cost efficiency by achieving up to a 70\% reduction in cost, underscoring its effectiveness in optimizing resource usage.

\section{Related Work}
\label{sec:related}

FL presents unique challenges when deployed in cloud environments, particularly in terms of cost efficiency, resource heterogeneity, and scheduling complexity \cite{kairouz2021advances}. This section reviews the existing literature pertinent to these challenges, contextualizing the contributions of our proposed system, FedCostAware. Specifically, we examine prior work in cloud cost optimization for model training, computing heterogeneity, and straggler management.

The computational demands of FL, especially when training large models among numerous clients, require cost optimization when using cloud resources. Although a lot of current research focuses on reducing communication overhead through techniques like model compression, sparsification, quantization, pruning, reduced synchronization frequency, or hierarchical aggregation \cite{wilkins2024fedsz,bai2024fedspallm,singhal2025fed}, the computational cost, particularly GPU-hours in cloud computing settings, is also a significant factor that has been overlooked until now. Cloud providers such as Amazon Web Services (AWS) and Google Cloud Platform (GCP) offer spot instances or preemptible instances as a lower-cost alternative to on-demand computing resources, allowing users to leverage spare compute capacity at dynamic prices. However, the volatility of spot instances due to potential preemptions poses a challenge. Effective usage of spot instances requires fault tolerance, often through checkpointing, a mechanism compatible with the iterative nature of FL. Previous work, such as spotDNN \cite{shang2023spotdnn},  aims to use spot instances for predictable Deep Neural Network (DNN) training performance by modeling resource bottlenecks and relying on asynchronous parallelism or takeover instances to handle preemptions. Other systems like Spotnik \cite{wagenlander2020spotnik} adapt communication collectives, Cynthia \cite{zheng2019cynthia} uses performance modeling for provisioning, while others leverage Markov Decision Process for better state prediction \cite{qiu2025convex}. FedCostAware differs by specifically targeting synchronous FL, reacting to real-time spot prices across diverse cloud regions/zones, and actively managing instance lifecycles during the inherent idle periods of synchronous protocols to minimize cost, rather than focusing solely on performance or relying on asynchronous methods.

Heterogeneity among FL clients is another key factor that impacts the cost-efficiency of FL. FL clients may have varying amounts of training data and differing computational capabilities, leading to significant variance in local training times. In synchronous FL algorithms, where the server has to wait for all clients to complete their local training before updating the global model, substantial computing resources can be wasted while waiting for slower clients (stragglers). This inefficiency becomes especially costly when relying on external cloud compute providers.  Common approaches in the literature include selective participation of clients based on the system or statistical utility, as implemented in systems such as Oort \cite{lai2021oort} and HACCS \cite{wolfrath2022haccs}. Other techniques focus on adapting client computational workloads within each local training round based on client capabilities through methods such as FedProx \cite{li2020federated}, HeteroFL \cite{diao2020heterofl}, and FedCompass \cite{li2024fedcompass}, or employing algorithmic modifications such as the layer-wise gradient aggregation found in Straggler-Aware Layer-wise Federated Learning (SALF) \cite{lang2024stragglers}. However, these techniques often involve trade-offs in terms of model convergence, fairness, or algorithm complexity. Alternatively, asynchronous FL protocols, such as FedAsync \cite{xie2019asynchronous}, FedBuff \cite{nguyen2022federated}, and AREA \cite{iakovidou2024asynchronous}, mitigate straggler delays by updating the global model and returning it to the client immediately upon receiving each client’s model update. However, these methods introduce staleness in updates, potentially compromising model accuracy and stability, which is undesirable in critical applications. In contrast, our work, FedCostAware, proposes a system-level solution within the synchronous paradigm. Instead of altering the core FL algorithm or discarding data from any client, it directly mitigates the financial impact of straggler delays by intelligently managing cloud spot instance lifecycles, terminating them during inevitable idle periods and proactively restarting them, ``pre-warming'' to maintain computational efficiency while preserving the integrity and convergence properties of standard synchronous aggregation.

\section{Algorithm Overview}
\label{sec:algorithm}
\begin{figure}[htbp]
    \centering
    \includegraphics[width=\linewidth]{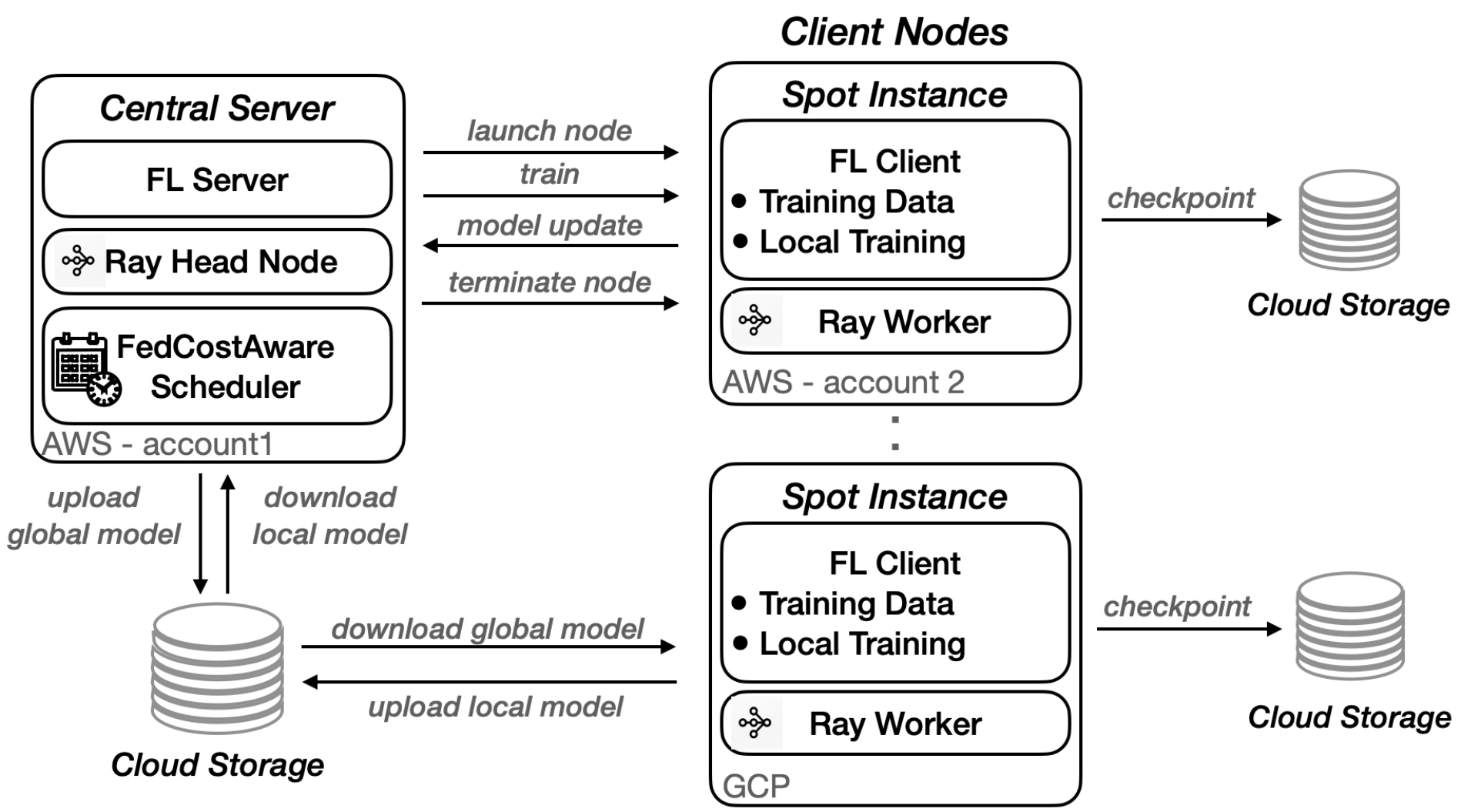}
    \caption{\textbf{System Architecture of the Cost-Aware Federated Learning System.} The central server coordinates the process with the FL server for model aggregation, the Ray head node for cluster management, and the FedCostAware scheduler for cost optimization. The central server launches/terminates client instances, initiates training, receives updates, and handles model transfers to/from the cloud storage. Client nodes are spot instances across different cloud providers. Each has an FL client for local training and a Ray worker for tasks, and checkpoints the model state to the cloud storage for fault tolerance. The server's cloud storage is used for data transfer between the server and client nodes.}
    \label{fig:architecture}
\end{figure}

Our proposed cost-aware scheduling algorithm operates within a distributed cloud environment, orchestrating synchronous FL across clients with heterogeneous resources and different budgets, leveraging spot instances for cost savings. This section details the system architecture, the core algorithmic workflow, dynamic instance management strategies, and mechanisms for fault tolerance and budget adherence.

The scheduler of FedCostAware is implemented by extending \text{Ray}'s \cite{moritz2018raydistributedframeworkemerging} cluster management capabilities---a framework designed for building and scaling distributed applications---to enable a single FL cluster to manage clients running on diverse cloud infrastructures (e.g., AWS, GCP, Azure). The design is inspired by previous works that use Ray to connect hybrid platforms~\cite{Ray-Cloud, Ray-Cloud-op}. As for FL infrastructure, FedCostAware builds upon the Advanced Privacy-Preserving Federated Learning (APPFL) framework \cite{ryu2022appfl, li2024advances}, which is an open-source software framework that enables researchers and developers to implement, test, and validate various privacy preserving FL techniques. This foundation makes our cost-aware functionalities readily adaptable and easily available for various FL workflows. FedCostAware minimizes GPU idle time and optimizes spot instance utilization, preserving the convergence benefits of synchronous training while substantially reducing operational costs and respecting heterogeneous financial capabilities, particularly within the resource-constrained environments often found in collaborative biomedical research.

\subsection{System Architecture and Environment Setup}
\label{subsec:system_architecture}

\begin{figure*}[t]
    \centering
    \includegraphics[width=0.9\textwidth]{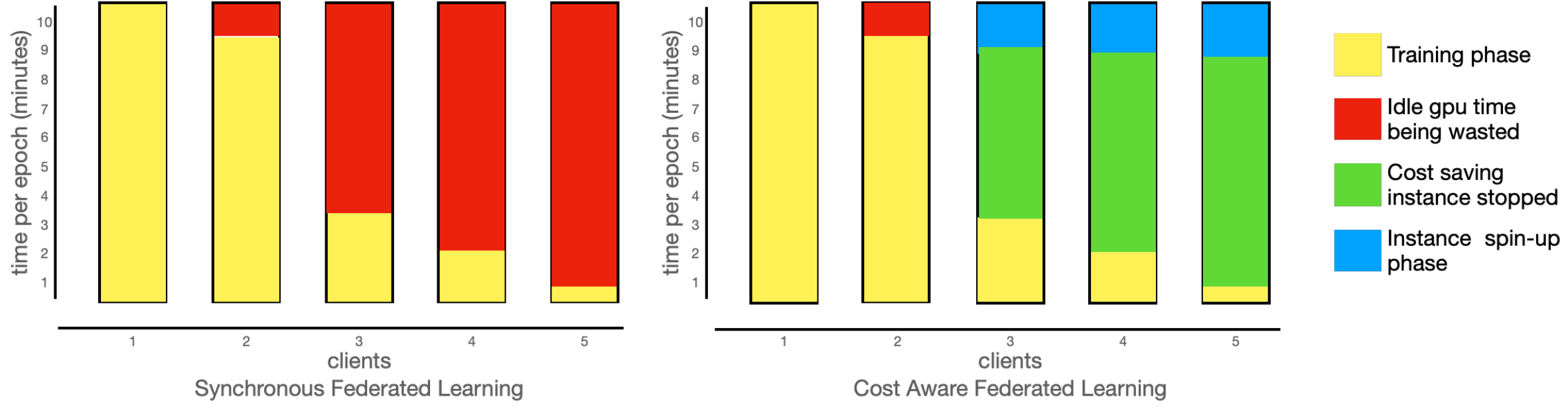}
    \caption{\textbf{Comparison of Client Activities per Epoch Between Standard and Cost-Aware Synchronous FL Processes.} In standard synchronous FL, significant GPU times are wasted for faster clients while waiting for the slowest client. The proposed cost-aware scheduling algorithm converts this idle time into cost savings by stopping instances and proactively restarting them just before the next training round begins, synchronized with the slowest client's completion.}
    \label{fig:intuition}
\end{figure*}

The system assumes a central server that coordinates the FL process as shown in Figure \ref{fig:architecture} and multiple clients who participate in training a model. Key configuration-related information needs to be provided to the system by the central server for cluster formation and execution orchestration.

\textbf{Configurations:} The system requires initial configuration details for each participating client, including their respective training budgets, cloud provider credentials (e.g., AWS, GCP) with necessary permissions for instance lifecycle management, and any FL-specific hyperparameters.

\textbf{Cluster Formation:}
\begin{itemize}
    \item \textit{Dynamic Resource Allocation:} Given the client configurations, the FedCostAware algorithm first queries real-time spot instance pricing across various cloud regions and availability zones compatible with the clients' potential locations. It then selects the most cost-effective options to minimize computing expenses during the cluster setup and subsequent training rounds.
    \item \textit{Ray Cluster Utilization:} The system leverages {Ray Cluster} capabilities to manage distributed resources. A {Ray Cluster} consists of a head node, which runs the central coordinating logic (the Ray head node), and multiple worker nodes where client computations are executed. Ray provides essential tools for distributed task execution and includes an autoscaler that can dynamically adjust the number of worker nodes based on workload demands, although our algorithm implements custom logic for more fine-grained and cost-aware scaling.
    \item \textit{Heterogeneous Cloud Support:} A key contribution of this work is a {custom node launcher} integrated with Ray, which enables the creation of a single and unified cluster that spans various cloud environments. For example, it allows one Ray cluster managed by the central server to include worker nodes running in different \textit{AWS} accounts, or even nodes running across different cloud providers (e.g., one client on \textit{AWS}, another on \textit{GCP}), provided the necessary credentials are supplied. The central Ray server acts as a unified control plane, managing FL training on behalf of all clients regardless of their underlying cloud infrastructure.
\end{itemize}

\textbf{Execution Orchestration:}
\begin{itemize}
    \item \textit{Ray Server:} The central Ray server orchestrates the entire FL experiment workflow. It dispatches tasks (as predefined methods), such as local model training, to be executed remotely on the clients' provisioned spot instances.
    \item \textit{Ray Actors:} Communication and state management for each client are handled using {Ray actors}. An actor in Ray is a stateful worker process or object. Unlike stateless Ray tasks, actors can maintain internal state across multiple method invocations. This makes them ideal for representing FL clients, allowing the server to invoke training or data-retrieval methods on a specific client actor, which can maintain its local model, data references, and state (like current budget usage) between different local training rounds.
    \item \textit{Dynamic Scaling:} While Ray's native autoscaler provides basic scaling, our algorithm employs {Ray actors} and custom logic to manage node lifecycles more proactively based on cost, budget, and FL synchronicity requirements (detailed in Section~\ref{subsec:dynamic_instance_management}). Nodes are scaled up when needed for a task and actively scaled down (terminated) during anticipated idle periods.
\end{itemize}

\subsection{Cost-Aware Scheduling and Execution Workflow}
\label{subsec:workflow}

Figure~\ref{fig:intuition} presents the intuition behind the design of our cost-aware FL scheduling algorithm: terminating the instances for clients finishing early and proactively restarting them right before the next training round begins. The algorithm proceeds through distinct phases elaborated below, incorporating cost efficiency throughout the synchronous FL process.

\textbf{Initialization:} The system begins by loading all client configurations, including budgets and credentials. Continuous budget tracking is initiated for each client via a background monitoring process.

\textbf{Calibration Phase:}
This phase establishes baseline estimates for different execution times of epochs as well as the instance spin-up time before dynamic optimizations begin. The rationale for estimating different execution epoch times is that the first epoch on a newly started instance might perform relatively longer than subsequent epochs. Below lists the times to estimate at the beginning of the FL process.
\begin{itemize}
    \item \textit{Round 1 (Cold Start):} The system launches spot instances for clients and sends the first training epoch task. The total time measured for each client in this round includes the first epoch's execution time just after an instance spins up, providing the initial estimate for (\textit{$T_{epoch\_cold}$}).
    \item \textit{Round 2 (Warm Start):} Immediately following Round 1, without terminating the instances, the system sends the second training epoch task. The time measured in this round primarily reflects the execution time on an already running ('warm') instance, providing the initial estimate for (\textit{$T_{epoch\_warm}$}).
    \item \textit{Initial Spin-up Estimate:} The initial spin-up time (\textit{$T_{est\_spin\_up}$}) can be estimated based on the total time observed in Round 1 relative to the execution time.
    \item \textit{Commencement of Optimization:} The dynamic instance termination logic begins operation only after these initial two calibration rounds, using the gathered estimates (\textit{$T_{epoch\_cold}$}, \textit{$T_{epoch\_warm}$}, \textit{$T_{est\_spin\_up}$}) to predict task completion times for lifecycle management.
\end{itemize}

\textbf{Synchronous Training Loop:} The core FL process begins:
\begin{itemize}
    \item The server sends the training task to all selected clients, instructing them to perform local training for one full epoch.
    \item The server waits to receive the resulting model updates from all participating clients in the current round for central aggregation.
\end{itemize}

\textbf{Dynamic Estimation Updates:} Whenever a resulting model update is received from a client:
\begin{itemize}
    \item The estimated \textit{$T_{epoch\_cold}$} and \textit{$T_{epoch\_warm}$} for that client are updated using an Exponential Moving Average (EMA) to smooth out variations and adapt to potential changes in computation speed.
    \item If receiving the result requires spinning up a new spot instance (e.g., due to prior termination or interruption), the estimated \textit{$T_{est\_spin\_up}$} is also updated using EMA based on the observed time. If the instance was already running, the \textit{$T_{est\_spin\_up}$} estimate remains unchanged.
\end{itemize}

\subsection{Dynamic Instance Management for Cost Optimization}
\label{subsec:dynamic_instance_management}
\begin{lstlisting}[
    style=PythonStyle,
    label={lst:behavior},
    caption={Instance Termination and Pre-warming Logic.},
    float,
    floatplacement=t
]

# Arguments:
# client_i - ID of the client being evaluated
# F_i - Finish time of client_i
# C_round - Set of clients in the current round
# q_spinup - Prewarm scheduling queue
# params - Structure with:
#    StartTime: client start times
#    IsColdStart: cold start flags
#    T_epoch_cold: cold-start training times
#    T_epoch_warm: warm-start training times
#    T_spinup: spin-up durations
#    T_threshold: idle time threshold

def evaluate_termination(client_i, F_i, C_round, params, q_spinup):
    F_s = estimate_slowest_finish_time(C_round, params)
    idle_time = F_s - F_i
    T_spin_up = params.T_spinup[client_i]

    if idle_time - T_spin_up > params.T_threshold:
        send_terminate_signal(client_i)
        prewarm_start_time = (F_s - T_spin_up - 
                              T_buffer)
        add_to_prewarm_queue(client_i, prewarm_start_time, q_spinup)
        return True
    return False

def estimate_slowest_finish_time(C_round, params):
    est_finish_times = []
    for c in C_round:
        if params.IsColdStart[c]:
            est_time = (params.StartTime[c] +
                        params.T_spinup[c] +
                        params.T_epoch_cold[c])
        else:
            est_time = (params.StartTime[c] +
                        params.T_epoch_warm[c])
        est_finish_times.append(est_time)
    return max(est_finish_times)

def send_terminate_signal(client_i):
    # calls custom Ray API for termination

def add_to_prewarm_queue(client_i, prewarm_start_time):
    q_spinup[client_i] = prewarm_start_time
    # the scheduler manages spinup

\end{lstlisting}

At the core of FedCostAware is the dynamic management of spot instances to minimize costs associated with idle time during synchronous waits. It automatically and intelligently terminates idle clients and proactively restarts them based on the time estimations, with the logic shown in Listing~\ref{lst:behavior}.

\textbf{Instance Termination:} After receiving a result from a client (\textit{$client_i$}) but before the global aggregation step, which requires results from all clients, the scheduler evaluates whether to temporarily terminate \textit{$client_i$}'s spot instance according to the following logic.
\begin{itemize}
\item The scheduler estimates the completion time for each client in the current round, based on whether the client’s instance is ``cold'' (newly launched) or ``warm'' (already running), using the corresponding time estimates (\textit{$T_{epoch\_cold}$} or \textit{$T_{epoch\_warm}$}).
\item The scheduler then obtains the \text{slowest finish time} (\textit{$F_s$}) by taking the maximum of the individual estimated completion times.
\item When \textit{$client_i$} completes its task at its finish time (\textit{$F_i$}), the scheduler computes its potential idle time as: $\textit{$T_{idle}$} = \textit{$F_s$} - \textit{$F_i$}$.
\item If $\textit{$T_{idle}$} > \textit{$T_{est\_spin\_up}$}$ and the potential savings ($\textit{$T_{idle}$} - \textit{$T_{est\_spin\_up}$}$) exceeds a predefined threshold \textit{$T_{threshold}$}, a termination signal is sent to \textit{$client_i$}'s instance.
\item The scheduler uses a custom API we wrote in Ray's autoscaler to request the termination of specific nodes based on its cost-management policy.
\end{itemize}

\textbf{Proactive Spin-up (Pre-warming):} If a client's instance is terminated and additional local training rounds are expected, the scheduler adds the client to a {spin-up queue}. The target time to initiate the spin-up process for \textit{$client_i$} is calculated as: $\textit{spin\_up\_start\_time} = \textit{$F_s$} - \textit{$T_{est\_spin\_up}$}$, with an additional buffer to ensure the instance is up before the next round begins. This buffer prevents delays from affecting the start of the next round, as the algorithm is synchronous and requires all clients to be ready. This ``pre-warming'' step ensures the client's spot instance is requested and likely ready just before the next round of training is scheduled to begin, minimizing delays caused by instance boot-up while still saving costs during the idle period.

% \begin{algorithm}[H]
% \caption{Instance Termination and Pre-warming Logic}
% \label{alg:termination_logic_concise}
% \begin{minipage}{\linewidth}
% \small
% \begin{algorithmic}[1]
% \Procedure{EvaluateTermination}{$client_i$, $F_i$, $C_{round}$, $Params$, $Q_{spinup}$}
%     \State $F_s \gets \Call{EstimateSlowestFinishTime}{C_{round}, Params}$
%     \State $idle\_time \gets F_s - F_i$
%     \State $T_{est\_spin\_up} \gets Params.T_{spinup}[client_i]$

%     \If{$(idle\_time - T_{est\_spin\_up} > Params.T_{threshold})$}
%         \State \Call{SendTerminateSignal}{$client_i$}
%         \State $PrewarmStartTime \gets F_s - T_{est\_spin\_up}$
%         \State \Call{AddToPrewarmQueue}{$client_i$, $PrewarmStartTime$, $Q_{spinup}$}
%         \State \Return \textit{True}
%     \EndIf
%     \State \Return \textit{False}
% \EndProcedure

% \vspace{1em}

% \Function{EstimateSlowestFinishTime}{$C_{round}$, $Params$}
%     \State $EstFinishTimes \gets \emptyset$
%     \ForAll{$c \in C_{round}$}
%         \If{$Params.IsColdStart[c]$}
%             \State $EstFinishTime_c \gets Params.StartTime_r + Params.T_{spinup}[c] + Params.T_{epoch\_cold}[c]$
%         \Else
%             \State $EstFinishTime_c \gets Params.StartTime_r + Params.T_{epoch\_warm}[c]$
%         \EndIf
%         \State Add $EstFinishTime_c$ to $EstFinishTimes$
%     \EndFor
%     \State \Return $\max(EstFinishTimes)$
% \EndFunction
% \end{algorithmic}
% \end{minipage}
% \end{algorithm}

\subsection{Fault Tolerance}
\label{subsec:fault_tolerance_budget}
\begin{figure}[htbp]
    \centering
    \includegraphics[width=0.95\linewidth]{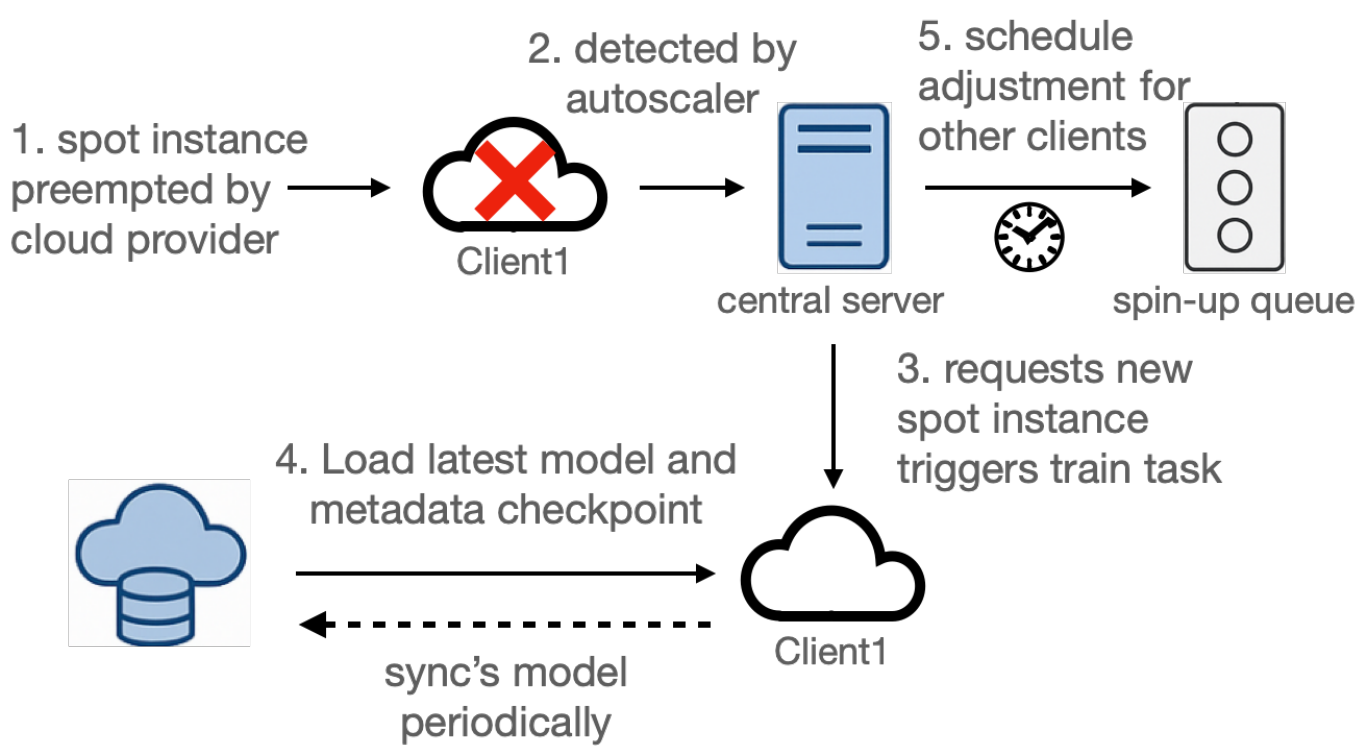}
    \caption{\textbf{The System's Fault Tolerance Mechanism.} Recovering from spot instance interruptions by checkpointing and schedule adjustment.}
    \label{fig:fault_tolerance}
\end{figure}

Figure~\ref{fig:fault_tolerance} presents the robust fault tolerance mechanisms incorporated into FedCostAware for handling the inherent volatility of spot instances. The details of the fault tolerance mechanisms are listed as follows.

\begin{itemize}
    \item \textit{Checkpointing and Recovery:} We implement a fault tolerance mechanism within the {APPFL} framework as shown in Figure \ref{fig:fault_tolerance}. During each round of local training, clients periodically checkpoint their local model state to persistent cloud storage (e.g., \textit{AWS S3}). If a spot instance is preempted (interrupted) by the cloud provider, upon restarting on a new instance, the central server identifies the last task the client was executing. The server then re-sends this task, prompting the client to load the latest checkpoint from cloud storage and resume training precisely from the saved step, thus minimizing lost computation.
    \item \textit{Dynamic Schedule Adjustment and Enhanced Cost Saving:} Given that spot instances can be preempted with little notice, our scheduler includes dynamic adjustments to further optimize cost savings. When an interruption occurs and a client restarts (loading from a checkpoint), the system estimates its new expected completion time for the current epoch (\textit{crashed\_client\_recovery\_finish\_time}). This allows the pre-warming schedule (spin-up queue) for all other temporarily terminated clients to be updated. Their \textit{spin\_up\_start\_time} is recalculated using: \textit{max(\textit{original\_slowest\_finish\_time}, \textit{crashed\_client\_recovery\_finish\_time}) - \textit{$T_{est\_spin\_up}$}}. This prevents premature spin-ups and idle time for other clients if the recovering client significantly delays the round's completion. Instead, they remain powered off, resulting in additional cost savings. Without such dynamic adjustments, a standard synchronous FL setup would waste GPU resources by keeping non-interrupted clients running while waiting for the affected client to recover. Our scheduler thus optimizes costs for the entire group during such events.

\end{itemize}

\subsection{Budget Adherence}
To ensure the whole FL training process adheres to the predefined budget set by each client, the scheduler checks each client's remaining budget against the estimated cost of participating in the upcoming round (based on the estimated epoch time and spot instance pricing) before initiating the tasks for each new FL round. If a client's remaining budget is deemed insufficient to complete the next round, the scheduler automatically excludes that client from participating in the current and all subsequent rounds. This ensures strict adherence to the predefined financial limits for each participant.

\section{Evaluations}
\label{sec:eval}
\begin{table*}[t]
    \centering
    \caption{Comparison of Experimental Costs and Savings Across Various Datasets and Scheduling Algorithms.}
    \label{tab:experiment_costs_savings}
    \sisetup{round-mode=places, round-precision=4} % Global setup, overridden by Savings column
    \begin{tabularx}{0.99\textwidth}{@{}l l l X S[table-format=5.0, round-precision=0] S[table-format=1.4, round-precision=4] S[table-format=2.4, round-precision=4] S[table-format=2.2, round-precision=2]@{}}
        \toprule
        \textbf{Dataset} & \textbf{\#Clients} & \textbf{\#Epochs} & \textbf{Algorithm} & {\textbf{Training Time (min)}} & {\textbf{Instance Cost (\$/hr)}} & {\textbf{Total Cost} $\downarrow$} & {\textbf{Savings (\%)} $\uparrow$} \\
        \midrule
        \multirow{3}{*}{Fed-ISIC2019}  & \multirow{3}{*}{6} & \multirow{3}{*}{20} & FedCostAware & 245 & 0.3951 & 7.17402825 & \multicolumn{1}{r}{\textbf{70.47\,\%}} \\ 
                                 &                    &                     & Sync FL Spot Instance & 243 & 0.3951 & 9.5238855 & 60.80357143\,\% \\ 
                                 &                    &                     & Sync FL On-demand & 243 & 1.0080 & 24.29784 & \multicolumn{1}{c}{---} \\
        \midrule
        \multirow{3}{*}{AI-READI}& \multirow{3}{*}{5} & \multirow{3}{*}{15} &  FedCostAware & 322 & 0.3946 & 8.330036333 & \multicolumn{1}{r}{\textbf{67.18\,\%}} \\ 
                                 &                    &                     & Sync FL Spot Instance & 319 & 0.3946 & 9.955027222 & 60.77693133\,\% \\ 
                                 &                    &                     & Sync FL On-demand & 319 & 1.0060 & 25.3805415 & \multicolumn{1}{c}{---} \\
        \midrule
        \multirow{3}{*}{CIFAR-10} & \multirow{3}{*}{4} & \multirow{3}{*}{20} &  FedCostAware & 396 & 0.3951 & 7.23987825 & \multicolumn{1}{r}{\textbf{72.22\,\%}} \\ 
                                 &                    &                     & Sync FL Spot Instance & 394 & 0.3951 & 10.2149593 & 60.80357143\,\% \\ 
                                 &                    &                     & Sync FL On-demand & 394 & 1.0080 & 26.060944 & \multicolumn{1}{c}{---} \\ 
        \midrule
        \multirow{3}{*}{MNIST}   & \multirow{3}{*}{3} & \multirow{3}{*}{10} &  FedCostAware & 95 & 0.3937 & 2.290134389 & \multicolumn{1}{r}{\textbf{67.04\,\%}} \\ 
                                 &                    &                     & Sync FL Spot Instance & 94 & 0.3937 & 2.7174105 & 60.89463496\,\% \\ 
                                 &                    &                     & Sync FL On-demand & 94 & 1.0060 & 6.948945 & \multicolumn{1}{c}{---} \\
        \bottomrule
    \end{tabularx}
\end{table*}

\begin{figure}[t]
    \centering
    \includegraphics[width=\linewidth]{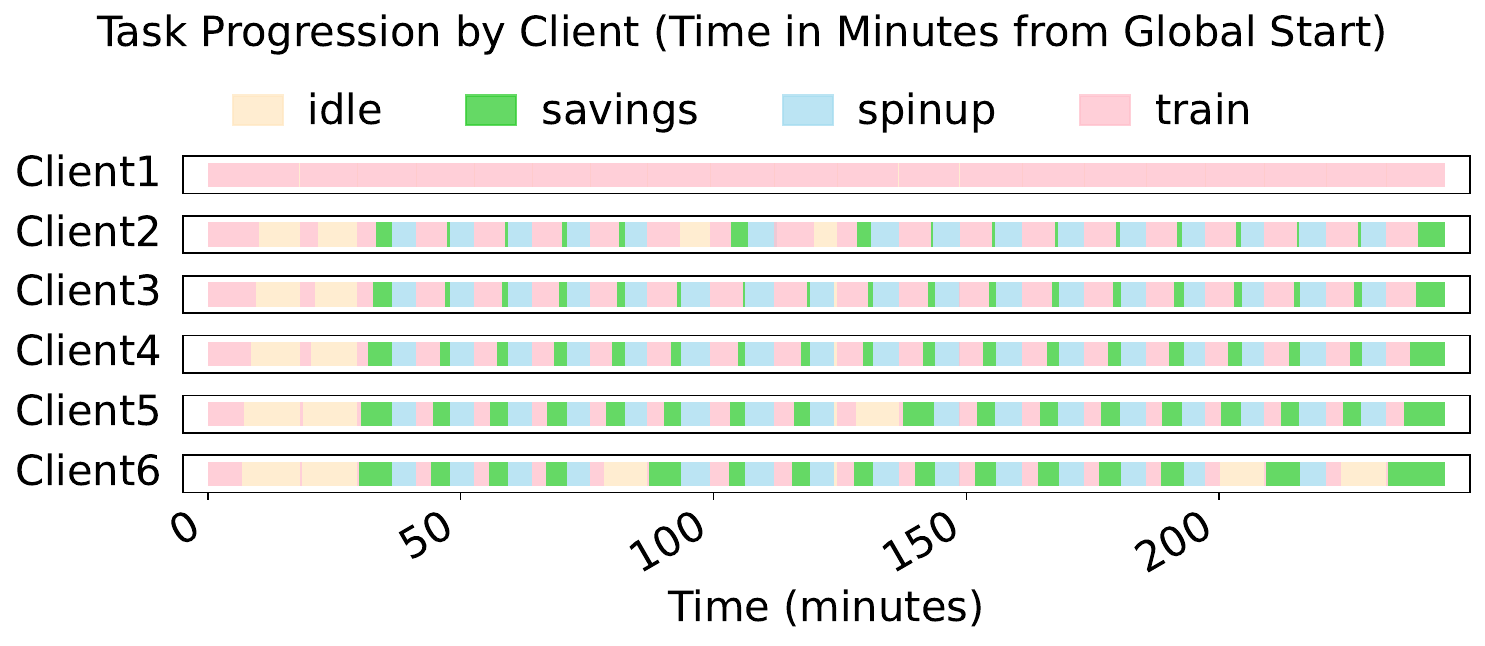}
    \caption{Client operational states over time for the Fed-ISIC2019 dataset, illustrating periods of training, spinup, idle time, and savings achieved for six clients over 20 epochs.}
    \label{fig:flamby_timeline}
\end{figure}

\begin{figure}[t]
    \centering
    \includegraphics[width=0.9\linewidth]{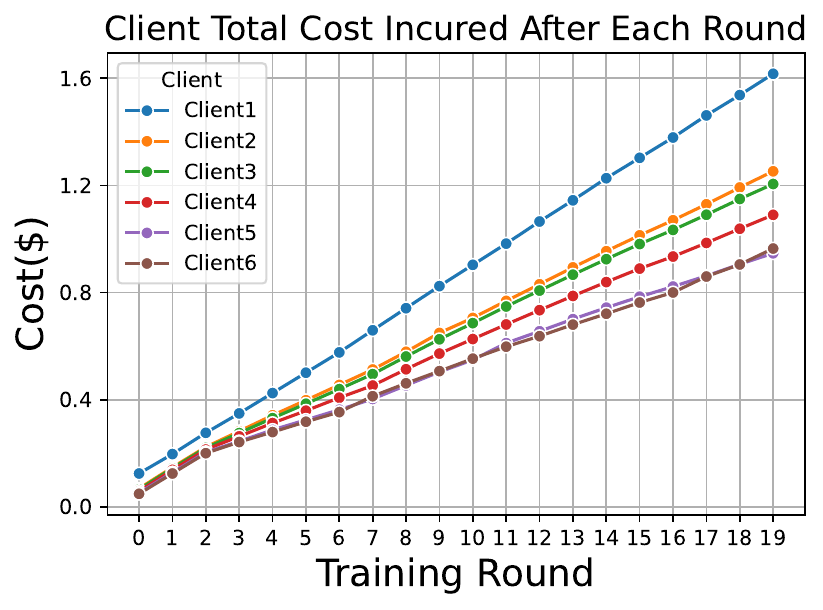}
    \caption{Client total cost incurred over 20 training rounds for the Fed-ISIC2019 dataset using the FedCostAware scheduler on g5.xlarge instances.}
    \label{fig:cost}
\end{figure}

\subsection{Experiment Settings}
\label{subsec:experiment_setup}

\textbf{Dataset:}
To evaluate our approach under varying degrees of statistical heterogeneity, we use both synthetically and naturally partitioned datasets. For synthetic benchmarks, we employ MNIST \cite{yann1998mnist}, CIFAR-10 \cite{krizhevsky2009learning}, and AI-READI \cite{AI-READI_Consortium_2024}, where the training goal for the AI-READI dataset is classifying retinal fundus photographs into different device categories based on their acquisition source. Each of these datasets is partitioned into several client splits using a dual Dirichlet method \cite{li2024fedcompass} to simulate non-IID heterogeneous data, modeling both class imbalance and variation in client data volume. To complement these, we use Fed-ISIC2019 from the FLamby benchmark \cite{ogier2022flamby}, a dataset involving melanoma classification from dermoscopy images sourced from six institutions. Unlike the others, this dataset maintains its natural, institution-based partitions, making it representative of real-world FL scenarios with inherent data silos.

\textbf{Cloud Settings:}
Experiments are run on AWS using the Ray framework for cluster management. GPU-enabled \textit{g5.xlarge} instances serve as client nodes, while a non-GPU \textit{t3.xlarge} instance acts as the central server. Model updates are exchanged through S3 using presigned URLs, although associated costs are negligible compared to those of EC2 usage. Importantly, all datasets remain within the secure cloud accounts of the respective clients. No raw data is transferred outside these accounts, and only model updates are exchanged, ensuring data locality and privacy.

\textbf{Simulation Setup:} Our simulation setup is tailored to evaluate the scheduling algorithm with realistic training durations of the clients, particularly for observing straggler effects. For MNIST and CIFAR-10, where the default training is rapid, we introduce client-specific scaling factors to per-epoch processing durations to simulate longer training time. For the more complex AI-READI and Fed-ISIC2019 datasets, inherent computational demands provide sufficiently long and variable per-epoch training times, making simulations unnecessary for scheduler testing.

\textbf{Training Settings:} Model selection is dataset-specific: for the naturally partitioned Fed-ISIC2019 dataset from FLamby, we use FLamby's default experimental settings and model architecture (EfficientNet \cite{tan2019efficientnet}) during the training. For the artificially partitioned MNIST dataset, we train a two-layer convolutional neural network on three simulated clients. As for the CIFAR-10 dataset, a ResNet-18 \cite{he2016deep} model is trained on four simulated clients. For the AI-READI dataset, a ResNet-50 model is trained on five simulated clients.

\subsection{Experiment Results}
\label{subsec:experiment_results}

The evaluation of our cost-aware scheduling algorithm, FedCostAware, demonstrates its effectiveness in reducing cloud computing costs for FL across multiple datasets. Table \ref{tab:experiment_costs_savings} presents a comparison of the total costs and savings achieved by FedCostAware compared to two baseline approaches: standard spot instances and on-demand instances. The results indicate that FedCostAware consistently outperforms both baselines by achieving significant cost reductions.

The most representative results come from the Fed-ISIC2019 and AI-READI datasets, both of which do not involve any artificially introduced wait time and reflect realistic FL workloads. For the Fed-ISIC2019 dataset, FedCostAware achieves a cost reduction of 70.47\% compared to on-demand instances. Similarly, for the AI-READI dataset, FedCostAware yields 67.18\% savings over on-demand instances. These results highlight FedCostAware’s real-world applicability, particularly in domains such as healthcare, where computational efficiency is crucial. In all datasets, FedCostAware consistently delivers the highest savings relative to the on-demand baseline, with a peak reduction of 72.22\% observed in the CIFAR-10 dataset. Even when compared to the more cost-effective spot instance baseline, FedCostAware achieves significant additional savings by reducing idle time and optimizing instance lifecycles.

Figure \ref{fig:flamby_timeline} provides an illustration of the operational states of six clients over time for the Fed-ISIC2019 dataset. In this example, the first client has the largest volume of data and consequently the longest local training time, so its compute instance remains active throughout the entire training process. In contrast, the other client instances are dynamically managed and intelligently terminated (shown as green savings) by FedCostAware to reduce machine idle time and optimize cost-efficiency. FedCostAware terminates the compute instances of clients that finish local training early and restarts them ahead of time to synchronize the whole training process. In addition, as the graph illustrates, the training time for clients after a cold start is relatively longer compared to the training time in a warm start scenario, highlighting the importance of accounting for both the cold start and the warm start training time during estimation. Figure~\ref{fig:cost} shows the accumulated costs for each client over 20 training rounds on the Fed-ISIC2019 dataset when using the FedCostAware algorithm. The plot highlights the cost variations across six clients due to heterogeneous resource usage and dynamic scheduling.

We also observed that using the exponential average to estimate spin-up time introduced at most $\sim$3 minutes of extra delay over the course of an entire dataset experiment, during which the slowest instance idled while waiting for others to restart for the next synchronous round.

No spot preemptions occurred during our experiments, even in long-running sessions exceeding six hours, suggesting that under certain conditions spot instances offer reliable, cost-efficient support for extended FL workloads when combined with fault tolerance. While our tests used \$1 GPU instances, the savings demonstrated by FedCostAware would scale substantially on higher-cost GPUs such as P4 (\$30/hr) or P5 (\$100/hr), where reducing idle time yields proportionally greater benefits.

\section{Conclusion and Future Work}
\label{sec:conclusion}
In this paper, we introduce FedCostAware, a cost-aware scheduling algorithm designed to optimize the efficiency of synchronous FL in cloud environments.  Our algorithm dynamically manages spot instance lifecycles, intelligently terminating instances during idle periods, and proactively initiating them before subsequent training rounds.  This approach minimizes GPU idle time and overall costs while adhering to heterogeneous client budgets.  Evaluations in various datasets demonstrate that FedCostAware significantly reduces expenses compared to the usage of standard spot instances and on-demand instances, achieving up to 72.22\% cost savings.  These results highlight the potential of FedCostAware to make FL more accessible and cost-effective, particularly for resource-constrained collaborative research in fields like biomedicine.

% In future work, our goal is to extend the capabilities of FedCostAware in several key directions. First, we will investigate its scalability with larger client populations, additional datasets, and more complex models, thereby broadening its applicability. Second, we will study the impact of diverse client participation strategies on model accuracy in budget-constrained environments, with the goal of improving budget utilization for better model performance. We also plan to adapt FedCostAware to support hybrid client deployments, enabling FL that integrates clients across both cloud platforms and high-performance computing (HPC) systems. This adaptation will facilitate the optimized use of HPC GPU resources and may further enhance both cost-efficiency and overall performance.

\section*{Acknowledgment}
This material is based upon work supported by the U.S. Department of Energy, Office of Science, under contract number DE-AC02-06CH11357. We also gratefully acknowledge Amazon Web Services for providing cloud computing credits that were used to assist with experiment efforts for this manuscript.

\bibliographystyle{IEEEtran}
\bibliography{references}

\begin{thebibliography}{10}
\providecommand{\url}[1]{#1}
\csname url@samestyle\endcsname
\providecommand{\newblock}{\relax}
\providecommand{\bibinfo}[2]{#2}
\providecommand{\BIBentrySTDinterwordspacing}{\spaceskip=0pt\relax}
\providecommand{\BIBentryALTinterwordstretchfactor}{4}
\providecommand{\BIBentryALTinterwordspacing}{\spaceskip=\fontdimen2\font plus
\BIBentryALTinterwordstretchfactor\fontdimen3\font minus \fontdimen4\font\relax}
\providecommand{\BIBforeignlanguage}[2]{{%
\expandafter\ifx\csname l@#1\endcsname\relax
\typeout{** WARNING: IEEEtran.bst: No hyphenation pattern has been}%
\typeout{** loaded for the language `#1'. Using the pattern for}%
\typeout{** the default language instead.}%
\else
\language=\csname l@#1\endcsname
\fi
#2}}
\providecommand{\BIBdecl}{\relax}
\BIBdecl

\bibitem{mcmahan2017communication}
B.~McMahan, E.~Moore, D.~Ramage, S.~Hampson, and B.~A. y~Arcas, ``Communication-efficient learning of deep networks from decentralized data,'' \emph{Artificial intelligence and statistics}, pp. 1273--1282, 2017.

\bibitem{li2020federated}
T.~Li, A.~K. Sahu, A.~Talwalkar, and V.~Smith, ``Federated learning: Challenges, methods, and future directions,'' \emph{IEEE Signal Processing Magazine}, vol.~37, no.~3, pp. 50--60, 2020.

\bibitem{kim2024privacy}
K.~Kim, K.~Raghavan, O.~Kotevska, M.~Dorier, R.~Madduri, M.~Ryu, T.~Munson, R.~Ross, T.~Flynn, A.~Kagawa \emph{et~al.}, ``Privacy-preserving federated learning for science: Challenges and research directions,'' in \emph{2024 IEEE International Conference on Big Data (BigData)}.\hskip 1em plus 0.5em minus 0.4em\relax IEEE, 2024, pp. 7849--7853.

\bibitem{fl_medimg}
G.~Kaissis, A.~Ziller, J.~Passerat-Palmbach, T.~Ryffel, D.~Usynin, A.~Trask, I.~Lima~Jr, J.~Mancuso, F.~Jungmann, M.-M. Steinborn \emph{et~al.}, ``End-to-end privacy preserving deep learning on multi-institutional medical imaging,'' \emph{Nature Machine Intelligence}, vol.~3, no.~6, pp. 473--484, 2021.

\bibitem{fl_cancer}
S.~Pati, U.~Baid, B.~Edwards, M.~Sheller, S.-H. Wang, G.~A. Reina, P.~Foley, A.~Gruzdev, D.~Karkada, C.~Davatzikos \emph{et~al.}, ``Federated learning enables big data for rare cancer boundary detection,'' \emph{Nature Communications}, vol.~13, no.~1, p. 7346, 2022.

\bibitem{hoang2023enabling}
T.-H. Hoang, J.~Fuhrman, R.~Madduri, M.~Li, P.~Chaturvedi, Z.~Li, K.~Kim, M.~Ryu, R.~Chard, E.~Huerta \emph{et~al.}, ``Enabling end-to-end secure federated learning in biomedical research on heterogeneous computing environments with {APPFLx},'' \emph{arXiv preprint arXiv:2312.08701}, 2023.

\bibitem{pfitzner2021federated}
B.~Pfitzner, N.~Steckhan, and B.~Arnrich, ``Federated learning in a medical context: a systematic literature review,'' \emph{ACM Transactions on Internet Technology (TOIT)}, vol.~21, no.~2, pp. 1--31, 2021.

\bibitem{ye2023heterogeneous}
M.~Ye, X.~Fang, B.~Du, P.~C. Yuen, and D.~Tao, ``Heterogeneous federated learning: State-of-the-art and research challenges,'' \emph{ACM Computing Surveys}, vol.~56, no.~3, pp. 1--44, 2023.

\bibitem{xie2019asynchronous}
C.~Xie, S.~Koyejo, and I.~Gupta, ``Asynchronous federated optimization,'' \emph{arXiv preprint arXiv:1903.03934}, 2019.

\bibitem{nguyen2022federated}
J.~Nguyen, K.~Malik, H.~Zhan, A.~Yousefpour, M.~Rabbat, M.~Malek, and D.~Huba, ``Federated learning with buffered asynchronous aggregation,'' in \emph{International conference on artificial intelligence and statistics}.\hskip 1em plus 0.5em minus 0.4em\relax PMLR, 2022, pp. 3581--3607.

\bibitem{li2024fedcompass}
\BIBentryALTinterwordspacing
Z.~Li, P.~Chaturvedi, S.~He, H.~Chen, G.~Singh, V.~Kindratenko, E.~A. Huerta, K.~Kim, and R.~Madduri, ``{FedCompass}: Efficient cross-silo federated learning on heterogeneous client devices using a computing power-aware scheduler,'' in \emph{The Twelfth International Conference on Learning Representations}, 2024. [Online]. Available: \url{https://openreview.net/forum?id=msXxrttLOi}
\BIBentrySTDinterwordspacing

\bibitem{iakovidou2024asynchronous}
C.~Iakovidou and K.~Kim, ``Asynchronous federated stochastic optimization for heterogeneous objectives under arbitrary delays,'' \emph{arXiv preprint arXiv:2405.10123}, 2024.

\bibitem{wilhelmi2022analysisevaluationsynchronousasynchronous}
\BIBentryALTinterwordspacing
F.~Wilhelmi, L.~Giupponi, and P.~Dini, ``Analysis and evaluation of synchronous and asynchronous flchain,'' 2022. [Online]. Available: \url{https://arxiv.org/abs/2112.07938}
\BIBentrySTDinterwordspacing

\bibitem{dun2023efficient}
C.~Dun, M.~Hipolito, C.~Jermaine, D.~Dimitriadis, and A.~Kyrillidis, ``Efficient and light-weight federated learning via asynchronous distributed dropout,'' in \emph{International Conference on Artificial Intelligence and Statistics}.\hskip 1em plus 0.5em minus 0.4em\relax PMLR, 2023, pp. 6630--6660.

\bibitem{kairouz2021advances}
P.~Kairouz, H.~B. McMahan, B.~Avent, A.~Bellet, M.~Bennis, A.~N. Bhagoji, K.~Bonawitz, Z.~Charles, G.~Cormode, R.~Cummings \emph{et~al.}, ``Advances and open problems in federated learning,'' \emph{Foundations and trends{\textregistered} in machine learning}, vol.~14, no. 1--2, pp. 1--210, 2021.

\bibitem{wilkins2024fedsz}
G.~Wilkins, S.~Di, J.~C. Calhoun, Z.~Li, K.~Kim, R.~Underwood, R.~Mortier, and F.~Cappello, ``{FedSZ}: Leveraging error-bounded lossy compression for federated learning communications,'' in \emph{2024 IEEE 44th International Conference on Distributed Computing Systems (ICDCS)}.\hskip 1em plus 0.5em minus 0.4em\relax IEEE, 2024, pp. 577--588.

\bibitem{bai2024fedspallm}
G.~Bai, Y.~Li, Z.~Li, L.~Zhao, and K.~Kim, ``{FedSpaLLM}: Federated pruning of large language models,'' \emph{arXiv preprint arXiv:2410.14852}, 2024.

\bibitem{singhal2025fed}
R.~Singhal, K.~Ponkshe, R.~Vartak, L.~R. Varshney, and P.~Vepakomma, ``{Fed-SB}: A silver bullet for extreme communication efficiency and performance in (private) federated lora fine-tuning,'' \emph{arXiv preprint arXiv:2502.15436}, 2025.

\bibitem{shang2023spotdnn}
R.~Shang, F.~Xu, Z.~Bai, L.~Chen, Z.~Zhou, and F.~Liu, ``{SpotDNN}: Provisioning spot instances for predictable distributed {DNN} training in the cloud,'' in \emph{2023 IEEE/ACM 31st International Symposium on Quality of Service (IWQoS)}.\hskip 1em plus 0.5em minus 0.4em\relax IEEE, 2023, pp. 1--10.

\bibitem{wagenlander2020spotnik}
M.~Wagenl{\"a}nder, L.~Mai, G.~Li, and P.~Pietzuch, ``Spotnik: Designing distributed machine learning for transient cloud resources,'' in \emph{12th USENIX Workshop on Hot Topics in Cloud Computing (HotCloud 20)}, 2020.

\bibitem{zheng2019cynthia}
H.~Zheng, F.~Xu, L.~Chen, Z.~Zhou, and F.~Liu, ``Cynthia: Cost-efficient cloud resource provisioning for predictable distributed deep neural network training,'' in \emph{Proceedings of the 48th International Conference on Parallel Processing}, 2019, pp. 1--11.

\bibitem{qiu2025convex}
S.~Qiu, H.~Wang, Y.~Zhang, Z.~Ke, and Z.~Li, ``Convex optimization of markov decision processes based on z transform: A theoretical framework for two-space decomposition and linear programming reconstruction,'' \emph{Mathematics}, vol.~13, no.~11, p. 1765, 2025.

\bibitem{lai2021oort}
F.~Lai, X.~Zhu, H.~V. Madhyastha, and M.~Chowdhury, ``Oort: Efficient federated learning via guided participant selection,'' in \emph{15th $\{$USENIX$\}$ Symposium on Operating Systems Design and Implementation ($\{$OSDI$\}$ 21)}, 2021, pp. 19--35.

\bibitem{wolfrath2022haccs}
J.~Wolfrath, N.~Sreekumar, D.~Kumar, Y.~Wang, and A.~Chandra, ``Haccs: Heterogeneity-aware clustered client selection for accelerated federated learning,'' in \emph{2022 IEEE international parallel and distributed processing symposium (IPDPS)}.\hskip 1em plus 0.5em minus 0.4em\relax IEEE, 2022, pp. 985--995.

\bibitem{diao2020heterofl}
E.~Diao, J.~Ding, and V.~Tarokh, ``{HeteroFL}: Computation and communication efficient federated learning for heterogeneous clients,'' \emph{arXiv preprint arXiv:2010.01264}, 2020.

\bibitem{lang2024stragglers}
N.~Lang, A.~Cohen, and N.~Shlezinger, ``Stragglers-aware low-latency synchronous federated learning via layer-wise model updates,'' \emph{IEEE Transactions on Communications}, 2024.

\bibitem{moritz2018raydistributedframeworkemerging}
\BIBentryALTinterwordspacing
P.~Moritz, R.~Nishihara, S.~Wang, A.~Tumanov, R.~Liaw, E.~Liang, M.~Elibol, Z.~Yang, W.~Paul, M.~I. Jordan, and I.~Stoica, ``Ray: A distributed framework for emerging {AI} applications,'' 2018. [Online]. Available: \url{https://arxiv.org/abs/1712.05889}
\BIBentrySTDinterwordspacing

\bibitem{Ray-Cloud}
\BIBentryALTinterwordspacing
T.~Liu, M.~Ellis, C.~Costa, C.~Misale, S.~Kokkila-Schumacher, J.~Jung, G.-J. Nam, and V.~Kindratenko, ``Cloud-bursting and autoscaling for python-native scientific workflows using ray,'' in \emph{High Performance Computing: ISC High Performance 2023 International Workshops, Hamburg, Germany, May 21–25, 2023, Revised Selected Papers}.\hskip 1em plus 0.5em minus 0.4em\relax Berlin, Heidelberg: Springer-Verlag, 2023, p. 207–220. [Online]. Available: \url{https://doi.org/10.1007/978-3-031-40843-4_16}
\BIBentrySTDinterwordspacing

\bibitem{Ray-Cloud-op}
\BIBentryALTinterwordspacing
T.~Liu, H.~Tao, Y.~Lu, Z.~Zhu, M.~Ellis, S.~Kokkila-Schumacher, and V.~Kindratenko, ``Automated data management and learning-based scheduling for {Ray}-based hybrid {HPC}-cloud systems,'' in \emph{Euro-Par 2024: Parallel Processing: 30th European Conference on Parallel and Distributed Processing, Madrid, Spain, August 26–30, 2024, Proceedings, Part I}.\hskip 1em plus 0.5em minus 0.4em\relax Berlin, Heidelberg: Springer-Verlag, 2024, p. 180–194. [Online]. Available: \url{https://doi.org/10.1007/978-3-031-69577-3_13}
\BIBentrySTDinterwordspacing

\bibitem{ryu2022appfl}
M.~Ryu, Y.~Kim, K.~Kim, and R.~K. Madduri, ``{APPFL}: open-source software framework for privacy-preserving federated learning,'' in \emph{2022 IEEE International Parallel and Distributed Processing Symposium Workshops (IPDPSW)}.\hskip 1em plus 0.5em minus 0.4em\relax IEEE, 2022, pp. 1074--1083.

\bibitem{li2024advances}
Z.~Li, S.~He, Z.~Yang, M.~Ryu, K.~Kim, and R.~Madduri, ``Advances in {APPFL}: A comprehensive and extensible federated learning framework,'' \emph{arXiv preprint arXiv:2409.11585}, 2024.

\bibitem{yann1998mnist}
L.~Yann, ``The mnist database of handwritten digits,'' \emph{R}, 1998.

\bibitem{krizhevsky2009learning}
A.~Krizhevsky, G.~Hinton \emph{et~al.}, ``Learning multiple layers of features from tiny images,'' 2009.

\bibitem{AI-READI_Consortium_2024}
\BIBentryALTinterwordspacing
A.-R. Consortium, ``Flagship dataset of type 2 diabetes from the {AI-READI} project (1.0.0),'' 2024. [Online]. Available: \url{https://doi.org/10.60775/fairhub.1}
\BIBentrySTDinterwordspacing

\bibitem{ogier2022flamby}
J.~Ogier~du Terrail, S.-S. Ayed, E.~Cyffers, F.~Grimberg, C.~He, R.~Loeb, P.~Mangold, T.~Marchand, O.~Marfoq, E.~Mushtaq \emph{et~al.}, ``Flamby: Datasets and benchmarks for cross-silo federated learning in realistic healthcare settings,'' \emph{Advances in Neural Information Processing Systems}, vol.~35, pp. 5315--5334, 2022.

\bibitem{tan2019efficientnet}
M.~Tan and Q.~Le, ``Efficientnet: Rethinking model scaling for convolutional neural networks,'' in \emph{International conference on machine learning}.\hskip 1em plus 0.5em minus 0.4em\relax PMLR, 2019, pp. 6105--6114.

\bibitem{he2016deep}
K.~He, X.~Zhang, S.~Ren, and J.~Sun, ``Deep residual learning for image recognition,'' in \emph{Proceedings of the IEEE conference on computer vision and pattern recognition}, 2016, pp. 770--778.

\end{thebibliography}

% \vspace{12pt}
% \color{red}
% IEEE conference templates contain guidance text for composing and formatting conference papers. Please ensure that all template text is removed from your conference paper prior to submission to the conference. Failure to remove the template text from your paper may result in your paper not being published.

\end{document}